# Structural and Magnetic Properties of $Co_{1-x}Fe_xSr_2YCu_2O_{7+\delta}$ compounds

Shiva Kumar Singh,[1,2*] Praveen Kumar,[1] M. Husain,[2] Hari Kishan[1] and V.P.S Awana[1,]

[1]National Physical Laboratory (CSIR), Dr. K.S. Krishnan marg, New Delhi-110012, India

[2]Department of Physics, Jamia Millia Islamia University, New Delhi-110025, India

## Abstract

Here we study the structural and magnetic properties of the $Co_{1-x}Fe_xSr_2YCu_2O_{7+\delta}$ compound ($0 \leq x \leq 1$). X-ray diffraction patterns and simulated data obtained from Rietveld refinement of the same indicate that the iron ion replacement in $Co_{1-x}Fe_xSr_2YCu_2O_{7+\delta}$ induces a change in crystal structure. The orthorhombic *Ima2* space group structure of Co-1212 changes to tetragonal *P4/mmm* with increasing Fe ($x \geq 0.5$) ion. The XPS studies reveal that both Co and Fe ions are in mixed states of 3+/4+ for the former and 2+/3+ in case of later. The magnetization with temperature follows Curie–Weiss behaviour, in the range 150-300 K and short magnetic correlations/spin glass like features below 150 K. The observed magnetic behaviour is due to competition of anti-ferro/ferromagnetic exchange interaction of $Co^{3+}$ (Intermediate spin: IS)-O-$Co^{3+}$ (IS)/$Co^{4+}$ (Low spin: LS) and $Fe^{3+}$ (High spin: HS)-O-$Fe^{2+}$ (LS)/ $Fe^{3+}$ (HS)/$Co^{3+}$ (IS)/$Co^{4+}$ (LS) states. Although none of the studied as synthesized samples in $Co_{1-x}Fe_xSr_2YCu_2O_{7+\delta}$ are superconducting, the interesting structural changes in terms of their crystallisation space groups and the weak magnetism highlights the rich solid state chemistry of this class of materials.

## 1. Introduction

Charge transport and high temperature superconductivity (HTSC) is believed to reside in the $CuO_2$ planes of all known HTSC cuprates, except that $CuO_{1+\delta}$ chains have been reported to participate in the *b*-axis transport of $YBa_2Cu_3O_{7-\delta}$ [1]. In $YBa_2Cu_3O_{7-\delta}$ ($CuBa_2YCu_2O_{7-\delta}$, Cu-1212) there are two different Cu sites, namely Cu1 and Cu2. Cu1 resides in $CuO_{1+\delta}$ chains and Cu2 in superconducting $CuO_2$ planes. Even at macroscopic level, any contravene in integral $CuO_2$ stacks, affects superconductivity drastically [2-3]. The $CuO_{1+\delta}$ chain acts as a charge reservoir and provides the mobile carriers to superconducting $CuO_2$ planes.

A variety of high-$T_c$ superconductive compounds are related with M-1212 structure. The M-1212 structure tolerates a wide range of single-element constituents such as Cu, Co, Fe, Nb, Ta, Ru, Hg, Tl, Al, Ga, and various cation mixtures as M [4-8]. Some of the M-1212 phases are well-established superconductors (viz. Cu-1212), whereas some of them are non-superconducting yet (viz. Nb/Al-1212). The M-1212 phases with M = Ga, Al and Co, particularly, have attracted considerable interest as potential superconducting candidates due to the complicated structure of their $MO_{1\pm\delta}$ charge reservoirs [4-8]. In the charge reservoir, the M cations are tetrahedrally coordinated by oxygen atoms to form chains of corner-sharing $MO_4$ tetrahedra that run diagonally relative to the perovskite base. Further, it has been reported that the $MO_4$ tetrahedra are arranged into two kinds of chain, L (left) and R (right), in which the tetrahedra rotate in different ways [7-9]. The M-1212 phases with M = Fe, Nb, Tl and Ru has a tendency to form $MO_6$ octahedra, which have similar rotation in each unit cell resulting in tetragonal structure. The oxygen stoichiometry plays a crucial role in determining superconductivity and structure of these compounds. Changes in the concentration of vacancies due to oxygen may lead to structural and electronic phase transitions. The polyhedra formation of charge reservoir blocks ($MO_{4/6}$) depends upon the oxygen intake ability of the M. It has been reported that in Fe-1212, Fe forms $FeO_6$ octahedra in oxygenated Fe-1212 system. But after annealing in nitrogen atmosphere, it loses oxygen and is left with $FeO_4$ tetrahedra [10] resulting tetragonal to orthorhombic structure of Fe-1212. The Co-1212 ($CoSr_2YCu_2O_{7+\delta}$) phase has orthorhombic structure with $CoO_4$ tetrahedra. However the Co-1212 phase with Ba ion on Sr ion site ($Cu_{1-x}Co_xBa_2YCu_2O_{7+\delta}$, x=0.84 composition) was reported to crystallize in tetragonal *P4/mmm* space group [11]. This suggests that besides M, other constituents are also the deciding factor in the structure formation of cuprates. M = Ga and Co are made superconducting after annealing in ultra high pressure oxygen. The structural changes and the reason behind superconductivity are not revealed yet. There may be ultra pressure oxygen introduces itself in reservoir blocks and causes change from tetrahedra (Co/GaO$_4$) to octahedra (Co/GaO$_6$) resulting Orthorhombic-Tetragonal: O-T transformation. We have taken Co (which prefers CoO$_4$ tetrahedra) and Fe (which prefers FeO$_6$ octahedra) formation in reservoir blocks to investigate structural changes. Also Cobalt and Iron both have ferromagnetic nature but the magnetic behaviour of their pervoskites compounds is quite different from each other. The distinct magnetic feature in these pervoskites is due to the various spin states of Co and Fe ions. There are reports on pervoskite cobaltites that spin states of cobalt can be low spin (LS) and mixture of intermediate spin (IS) and/or LS for tetravalent and trivalent cobalt ions respectively [12-16]. Actually the spin state of Co$^{3+}$ is controversial: the high spin state (HS, $t_{2g}^4 e_g^2$), intermediate spin state (IS, $t_{2g}^5 e_g^1$), and a superposition of HS and LS spin states are all proposed [16-18] in cobaltites. Here the magnetic nature of Co$_{1-x}$Fe$_x$-1212 whose structure belongs to HTSC cuprate family, can be explained by Goodenough-Kanamory rule of superexchange [19-21] as in cobaltites. Here we are revealing the structural changes and magnetic properties & effect of these properties on each other in charge transport.

**1. Experimental Details**

The samples are synthesized in air by solid-state reaction route. The stoichiometric mixture of Co$_3$O$_4$, Fe$_3$O$_4$, SrCO$_3$, Y$_2$O$_3$, and CuO are ground thoroughly, calcined at 900ºC for 12h and then pre-sintered at 950ºC and 980ºC for 15h with intermediate grindings. Finally, the powders are palletized and sintered at 1000ºC for 15h in air. The phase formation is checked for each sample with powder diffractometer, Rigaku (Cu-Kα radiation) at room temperature. The phase purity and lattice parameter refining are done by Rietveld refinement programme (Fullprof version).The magnetization measurements are carried out on Quantum Design SQUID magnetometer MPMS-XL. The samples have been characterized by X-ray photoelectron spectroscopy (XPS), working at a base pressure of 5x10$^{-10}$ torr. The chamber is equipped with a dual anode Mg-Kα (1253.6 eV) and Al-Kα (1486.6 eV) X-ray sources and a high-resolution hemispherical electron energy analyzer. We have used Mg-Kα X-ray source for our analysis. The calibration of the binding energy scale is done with the C 1s line 284.6 eV from the carbon contamination layer. The calibration of the binding energy scale has done with the C (1s) line at 284.6 eV. The core level spectra of Co and Fe have been deconvoluted in to the Gaussian components.

## 2. Results and Discussion

All the samples are crystallized in single phase which is confirmed from the Rietveld analysis of powder X-ray diffraction pattern. The compositions with x<0.5 are fitted in orthorhombic *Ima2* space group [Fig.1] whereas compositions with x≥0.5 are fitted in tetragonal *P4/mmm* space group [Fig.2]. The change in space group, from *Ima2* to *P4/mmm* appears in XRD pattern. The *020* peak associated with main *002* peak of *Ima2* space group disappears and single *103* peak of *P4/mmm* space group appears with increasing Fe ion concentration [Fig.3a]. This clearly suggests the absence of b-axis contribution i.e. the L and R chains, due to the rotation of CoO$_4$ tetrahedra in different direction within one unit cell, are disappearing. The same can be seen with 631 peak of *Ima2* space group [Fig.3b]. The lattice parameters obtained from Rietveld Refinement of the XRD shows that as Fe concentration increases on Co site there is a variable change in the lattice parameters [Table1]. The *a*-parameter increases from x=0.0 to x=0.3 composition. The *b*- and *c*-parameter almost remained constant from x=0.0 to x=0.3 composition. But for x=0.4 there is decrease in all these parameter. It can be interpreted as follows: Considering the ionic radii of Co ions {Co$^{3+}$ (CN=6) 0.545 Å LS, 0.56 Å IS, 0.61 Å HS; Co$^{4+}$ 0.40 Å, (CN=4), 0.53 Å HS (CN=6)}, Fe ions { Fe$^{2+}$ 0.61 Å LS (CN=6), Fe$^{3+}$ 0.49 Å (CN=4), 0.58 Å (CN=5), 0.55 Å LS, 0.645 Å HS (CN=6); Fe$^{4+}$ 0.585 Å (CN=6)} and Cu ions {Cu$^{2+}$ 0.57 Å (CN=4), 0.64 Å (CN=5), 0.73 Å (CN=6); Cu$^{3+}$ 0.54 Å LS (CN=6)} [22]. The change in lattice parameters from x=0.0 to 0.4 can be explained as: Co ions are in mixed 3+ (CN=6, IS) and 4+ (CN=4) [23] states are being replaced by Fe$^{2+/3+}$ ions. This is also evident from the XPS and M-T measurements of these compounds (to be discussed latter). The decrease in parameters for x=0.4 indicates towards origination of change in space group Orthorhombic (*Ima2*)-Tetragonal (*P4/mmm*): (O-T). This means that the FeO$_{4/6}$ tetra/octahedra are not as tilted as pure CoO$_4$ tetrahedra in x=0.0, but getting more similar rotation in consecutive unit cells leading decrease in lattice parameters. For compositions x≥0.5 the lattice parameters decreases with increasing iron concentration. There may be two reasons of this decrease, (1) The minor decrease in *a*- and *b*-parameters (in respect to concentration [see table]) is attributed to the same O-T (*Ima2*-*P4/mmm*) transition. As with decrease of Co concentration the CoO$_4$ tetrahedra density is also decreasing so there are more tetra/octahedrons rotated in centro symmetrical fashion, resulting shrinkage of *a*- and *b*-parameters. (2) The decrease in *c*-parameter (minor) can be attributed to the intermixing of Cu ions and Fe ions at Cu2 and Fe1 (Cu1) sites. There are reports that the Fe ion replaces Cu ion at Cu2 site up to 22%-47% [2-3]. Fe ions in all possible ionic (2+/3+/4+) and coordination states ($FeO_{4/6}$ tetra/octahedra) have larger ionic radii than Co ions so the lattice parameters should increase. Whereas Fe ion at Cu2 site i.e. in CN=5 have lower ionic radii than Cu ion in the same coordination but in different ionic state. However Cu ions replacing Co/Fe ions at Fe1(Cu1) site have larger ionic radii resulting increase in *c*-parameter. But this change is lower than former and threfore there is overall minor decrease in *c*-parameter. Thus it seems that at higher iron concentration Fe ion is replacing Cu ion at Cu2 site. Though X-ray cannot resolve issue of intermixing of Fe at Cu1 and Cu2 sites and exact percentage and only the neutron diffraction and Mossbauer spectroscopy can resolve[2,3,24]. But the same is clearly indicates towards intermixing of Cu ions and Fe ions at Cu2 and Fe1(Cu1) sites. The ambiguity in intermixing of Cu ions and Fe ions at Cu2 and Fe1(Cu1) sites in X–ray diffraction pattern of studied samples can be attributed to almost equal structure factors of iron and copper.

To find out the oxidation state of Co & Fe the XPS study have been carried out for two samples x=0.0 & 0.7 ($CoSr_2YCu_2O_{7+\delta}$ & $Co_{0.3}Fe_{0.7}Sr_2YCu_2O_{7+\delta}$). The Co (2p) & Fe (2p) core level spectra have been deconvoluted in to the different Gaussian component to find out the contribution of different ionic states. The deconvoluted Co (2p) core level spectra for samples x=0.0 & 0.7 are shown in Fig. 4a. Comparing to the spectrum of x=0.0, the binding energy of two main component of Co ($2p_{3/2}$ and Co $2p_{1/2}$), in the spectrum for the sample x=0.7, shifts towards higher binding energy. We have observed peak broadening in the x=0.7 in comparison to x=0.0. Deconvolution of Co (2p) core level spectra shows the presence of $Co^{3+}$ and $Co^{4+}$ with a satellite peak, at the binding energy of 779.45 eV & 781.85 eV for Co ($2p_{3/2}$) and 795.00 eV & 797.05 eV for Co($2p_{1/2}$) respectively in both compositions. The similar kinds of results have also been reported in literature regarding concentration and binding energy of $Co^{3+/4+}$ [23, 25]. Curve shows the domination of $Co^{3+}$ state over $Co^{4+}$ state for x=0.0 sample, but for x=0.7 both states are almost equal. This increase in $Co^{4+}$ ions concentration can be attributed due to the presence of major fraction of Fe in 2+ state in x=0.7. In x=0.7 composition there is slight increase in binding energy of both $Co^{3+}$ & $Co^{4+}$ component than that in x=0.0 composition. Fe ions are in mixed 2+/3+ state in x=0.7 composition as shown in Figure 4(b). The $Fe^{2+}$ ions are in higher concentration than $Fe^{3+}$ ions. The Fe $2p_{3/2}$ main peak maximum of the $Fe^{2+}$ component has a binding energy of 708.8 eV, while that of the $Fe^{3+}$ components is 710.6 eV as in γ-Fe$_2$O$_3$ reported in [26]. In γ-Fe$_2$O$_3$ iron ions coordinate both tetra and octahedrally with oxygen the same is in the studied samples. On the other hand, the 2$p$1/2 main peak has a binding energy of 721.66 eV and 723.16 eV for Fe$^{2+}$ and Fe$^{3+}$ ions respectively. The ionic composition of Co and Fe thus found by XPS study are supportive to magnetic behaviour of studied samples.

The magnetization measurement (M-T) (ZFC & FC) for all the samples is done in magnetic field of strength 100 Oe. The magnetic behaviour in x=0.0, 0.1, 0.2 & to some extent in 0.3 is like Curie-Weiss in the temperature range of 150-300 K [Fig 5a]. This is an intermediate behaviour of anti-ferromagnetic and ferromagnetic ordering. Below 150 K spin-glass or canted ferromagnetism type broad down-turn is observed in magnetization measurement of x=0.0, 0.1 & 0.2, which is dominated by the paramagnetic contribution below 50 K [27]. The magnetization behaviour of x= 0.4, 0.5 & 0.7 is slightly different from the x=0.0, 0.1, & 0.2. These concentrations show intermediate behaviour of antiferro/ferromagnetic nature in the temperature range 100-300 K in which ferromagnetic is dominating, as Fe ion concentration is increasing [Fig. 5b and 5c]. However, below 50 K the behaviour is same as that in lower concentrations. The x=1.0 composition is more prominent with ferromagnetic nature in which the paramagnetic to ferromagnetic transition occurs around 80 K in zero field cooled (ZFC) M-T plot. In Co$_{1-x}$Fe$_x$-1212 the presence of the Co$^{3+}$, Co$^{4+}$, Fe$^{2+}$ and Fe$^{3+}$ ions make the magnetic behaviour more complicated. Since low spin Co$^{3+}$ ions carry no magnetic moment [28] and high spin Co$^{3+}$ ions have greater ionic radii that are contradictory to Rietveld refined parameters so we can say Co$^{3+}$ ions in intermediate spin (IS) state. However the magnetic properties in the temperature range 50-300 K can be interpreted by Goodenough-Kanamory rule of superexchange which applies to interatomic spin-spin interactions between two atoms, each carrying a net spin, that are mediated by virtual electron transfers between the atoms (superexchange). This rule states that superexchange interactions are anti-ferromagnetic where the virtual electron transfer is between overlapping orbitals that are each half-filled, but they are ferromagnetic where the virtual electron transfer is from a half-filled to an empty orbital or from a filled to a half-filled orbital. In the lower concentration range (≤ 0.2) the weak anti-ferromagnetic behaviour (300-50K) can be explained by Co$^{3+}$(IS)-O-Co$^{3+}$(IS) and Co$^{3+}$(IS)-O-Co$^{4+}$(LS) electron exchange as in cobaltites [29-30]. Actually this rule cannot directly tell whether the exchange interactions through Co$^{3+}$(IS)-O-Co$^{3+}$(IS) and Co$^{3+}$(IS)-O-Co$^{4+}$(LS) (partially filled- partially filled and partially filled-empty e$_g$ orbitals) are ferromagnetic or anti-ferromagnetic because the sign of this interaction depends on the relative orientation of unoccupied/occupied $e_g$ orbitals. Although the relative orientation information of the orbitals in the present materials is not available but with the magnetic nature of studied samples and in the presence of variety of spins [$Co^{3+}$(IS), $Co^{4+}$(LS), $Fe^{2+}$(LS) and $Fe^{3+}$(HS)] we could propose that anti-ferromagnetic exchange and ferromagnetic exchange are competing here in which anti-ferromagnetic exchange is dominating. This results in weak anti-ferromagnetism and/or spin glass like behaviour since spins got frustrated with these competition. The magnetic behaviour of compounds with higher Fe ion concentrations ($0.4 \leq x \leq 1.0$) can be explained by the same exchange interaction. Here the $Fe^{3+}$(HS)-O-$Co^{3+}$(IS)/$Co^{4+}$(LS) and $Fe^{3+}$(HS)-O-$Fe^{2+}$(LS) and/or $Fe^{2+}$(LS)-O-$Cu^{3+}$(LS) and $Cu^{3+}$(LS)-O-$Co^{3+}$(IS)/$Co^{4+}$(LS) (in case of intermixing of Cu ions and Fe ions at Cu2 and Fe1 (Cu1) sites which infers from Rietveld refinement in compounds of higher Fe ion concentration) exchange interaction is being taken place. The ferromagnetic nature (ZFC) of x=1.0 also indicates about mixed state of $Fe^{2+/3+}$ ions since $Fe^{3+}$(HS)-O-$Fe^{2+}$(LS) exchange interaction is ferromagnetic. The same anti-ferromagnetic exchange and ferromagnetic exchange are in competition. But here the ferromagnetic exchange is dominating and is more prominent in higher Fe ion concentrations resulting weak ferromagnetism. However the contribution of ferromagnetic Fe spins towards domination of ferromagnetic nature in higher Fe ion concentration compounds cannot be excluded completely. In the lower temperature range (< 50 K) the frustration of spins due to these anti-ferro/ferromagnetic competition results in paramagnetic nature. The variation of moment is monotonic with the Fe concentration at 5 K and the same is non-monotonic at 300 K and it can be explained as follows: The magnitude of magnetization decreases from x=0.0 to x=0.3 in higher temperature range this is due to $Co^{3+}$ (IS) got replaced by $Fe^{2+}$ (LS)/ $Fe^{3+}$ (HS). Since $Fe^{2+}$ (LS) is in higher concentration and have lower magnetic moment than $Co^{3+}$ (IS) {having higher number of unpaired electrons than $Fe^{2+}$ (LS)}. The magnitude of magnetization remains almost equal for x= 0.3, 0.4 and 0.5. However there is increase in the magnitude of magnetization, from x=0.5 to x=1.0 and x=1.0 have slightly higher magnitude than that of the x= 0.0. It is due: (1) as we concluded that in higher Fe concentration samples there is increase in $Co^{4+}$ (LS) and there may be intermixing of Fe and Cu ions at Cu1/Cu2 site. Thus replacement of $Co^{3+}$ (IS) by $Co^{4+}$ (LS)/ $Fe^{2+}$ (LS) results in decrease in magnitude however it is got compensated by increase in magnitude by $Cu^{3+}$ (LS)/ $Fe^{3+}$ (HS) ions. Hence there is small increase in intermediate concentrations ($0.3 \leq x \leq 0.7$). (2) For x=1.0 $Cu^{3+}$ (LS)/ $Fe^{3+}$ (HS) have higher magnetic moment than $Co^{3+}$ (IS)/ $Co^{4+}$ (LS). In lower temperature (<50 K) range the paramagnetic ordering starts after weak anti-ferromagnetic/spin glass (SG) in lower Fe concentrations and after weak ferromagnetic ordering in higher Fe concentration range and hence is monotonic.

**Conclusion**

The studied compounds show that with increasing Fe concentration both structural and magnetic properties changes. The orthorhombic (*Ima2* space group) structure of Co-1212 crystallizes in tetragonal *P4/mmm* of Fe-1212. XPS study reveals that Co ions are in mixed $Co^{3+/4+}$ states and with increasing Fe ions on Co site $Co^{4+}$ concentration increases. Whereas Fe ions are in $Fe^{2+/3+}$ state and $Fe^{2+}$ state is dominating. The observed magnetic nature is explained by famous Goodenough-Kanamory rule of superexchange. It is concluded that anti-ferro/ferromagnetic competition is responsible for observed magnetic behavior.

**Acknowledgements**

The authors would like to thank DNPL Prof. R. C. Budhani for his constant support and encouragement. One of the authors Shiva Kumar would like to acknowledge CSIR, India for providing fellowships. V P S Awana is also thankful to Prof. E. Takayama Muromauchi for his visit to NIMS, Japan and to carry out the magnetization measurements.

**Figure Caption**

Table 1: Rietveld Refined lattice parameters and unit cell volume $Co_{1-x}Fe_xSr_2YCu_2O_{7+\delta}$ ($0.0 \leq x \leq 1.0$) compounds.

Fig. 1: Rietveld fitted XRD pattern of $Co_{1-x}Fe_xSr_2YCu_2O_{7+\delta}$ (x=0.0, 0.2 & 0.4) samples with space group *Ima2*.

Fig. 2: Rietveld fitted XRD pattern of $Co_{1-x}Fe_xSr_2YCu_2O_{7+\delta}$ (x=0.5, 0.7 &1.0) samples with space group *P4/mmm*.

Fig. 3: (a) X-ray pattern of main *002* peak of *Ima2* and *103* peak of *P4/mmm* space group. The change in space group, from *Ima2* to *P4/mmm* can be seen as the *020* peak associated with main *002* peak of *Ima2* disappearing and single *103* peak of *P4/mmm* is appearing with increasing Fe ion concentration. The arrows shows shift towards lower angle in 2θ as volume increases in (x≤ 0.4) in *Ima2* space group, and towards higher angle in 2θ as volume decreases in (0.5≤ x≤ 1.0) in *P4/mmm* space group. (b) X-ray pattern of main *613* peak of *Ima2* and *213* peak of *P4/mmm*. The change in space group, from *Ima2* to *P4/mmm* can be seen as the *631* peak associated with main *613* peak of *Ima2* disappearing and single *213* peak of *P4/mmm* space group is appearing with increasing Fe ion concentration. The shift towards higher angle in 2θ is more prominent with *213* peak.

Fig. 4: (a) The Co2*p* XPS spectra for the sample *x*=0.1 and 0.7. (b) The Fe2*p* XPS spectra for the sample *x*=0.7. The dashed line represents the experimental curve and the solid line represents the resultant of fitted curve.

Fig. 5: (a) 1/M vs T $Co_{1-x}Fe_x Sr_2YCu_2O_{7+\delta}$ (x=0.0, 0.1, 0.2 & 0.3) samples (b) 1/M vs T $Co_{1-x}Fe_x Sr_2YCu_2O_{7+\delta}$ (x=0.4, 0.5, 0.7 & 1.0) samples. (c) M-H of x=0.7 sample at room temperature.

Fig.6: Magnetization vs temperature (M-T) behaviour of $Co_{1-x}Fe_xSr_2YCu_2O_{7+\delta}$ ($0.1 \leq x \leq 0.7$) ; inset shows the (M-T) behaviour of x=0.0 & 1.0.

**Table 1**

| Co$_{1-x}$Fe$_x$-1212 | x=0.0 | x=0.1 | x=0.2 | x=0.3 | x=0.4 | x=0.5 | x=0.7 | x=1.0 |
|---|---|---|---|---|---|---|---|---|
| **a** (Å) | 22.78(2) | 22.79(7) | 22.80(6) | 22.81(2) | 22.79(1) | 3.83(3) | 3.82(9) | 3.82(1) |
| **b** (Å) | 5.45(1) | 5.45(3) | 5.45(2) | 5.44(8) | 5.43(7) | 3.83(3) | 3.82(9) | 3.82(1) |
| **c** (Å) | 5.40(9) | 5.41(0) | 5.41(1) | 5.41(2) | 5.41(3) | 11.38(8) | 11.37(8) | 11.36(1) |
| **V** (Å$^3$) | 671.74(8) | 672.52(4) | 672.80(3) | 672.63(0) | 670.73(7) | 167.29(5) | 166.79(1) | 165.88(1) |
| **R$_p$** | 2.11 | 2.52 | 2.48 | 2.48 | 2.54 | 2.74 | 2.92 | 2.45 |
| **R$_{wp}$** | 2.85 | 3.42 | 3.37 | 3.23 | 3.30 | 3.76 | 3.98 | 3.34 |
| **Chi$^2$** | 2.35 | 3.18 | 3.21 | 2.66 | 2.89 | 3.72 | 3.84 | 2.20 |

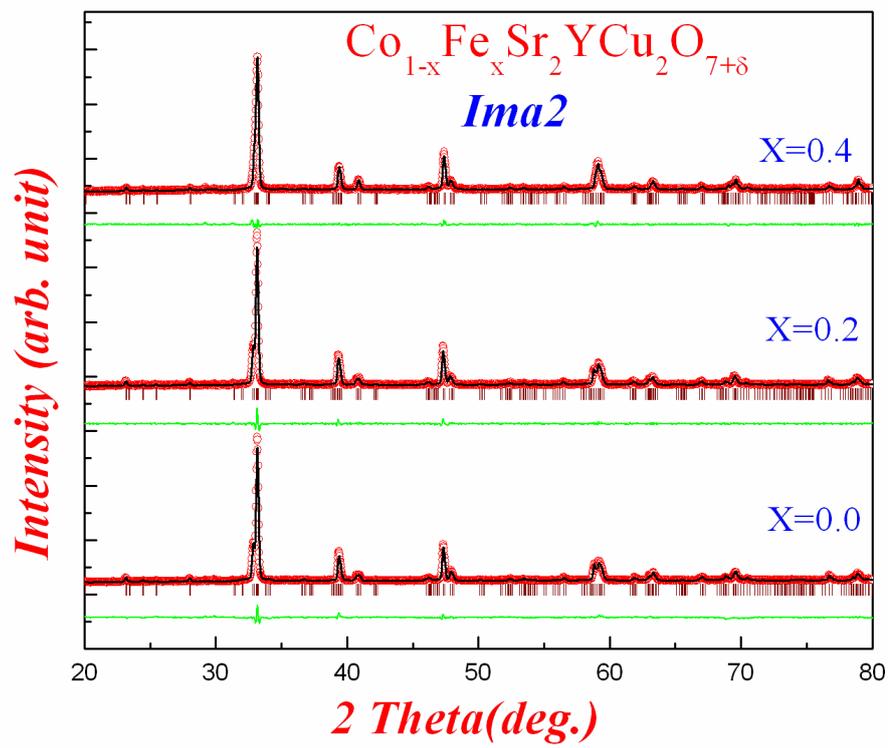

**Fig. 1**

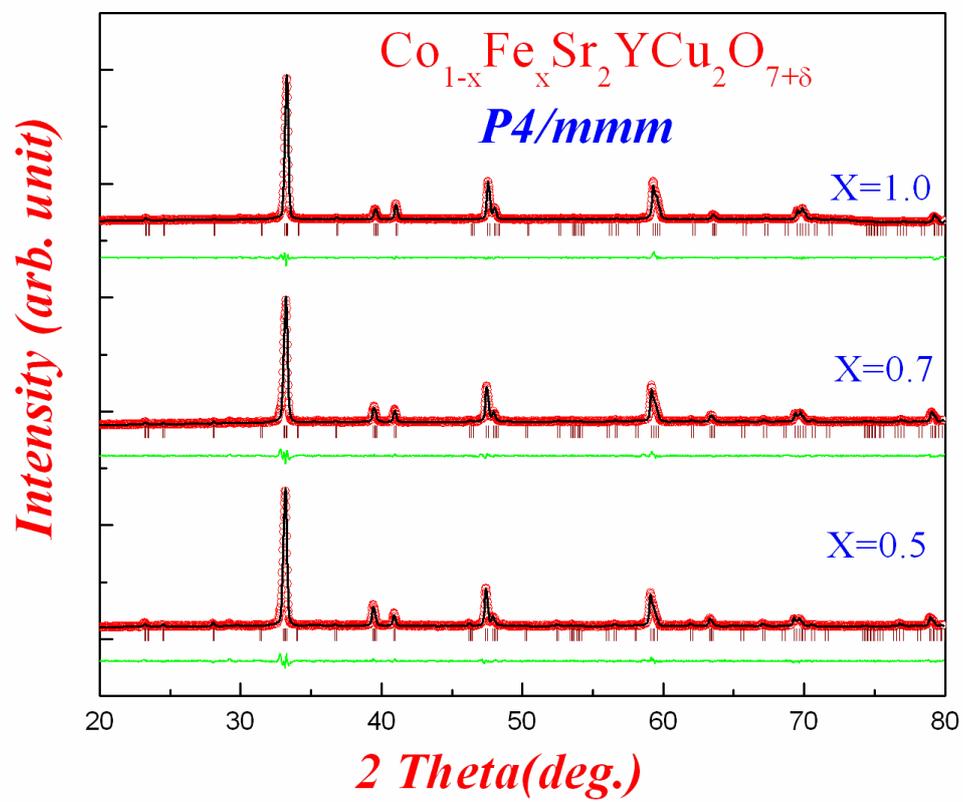

**Fig. 2**

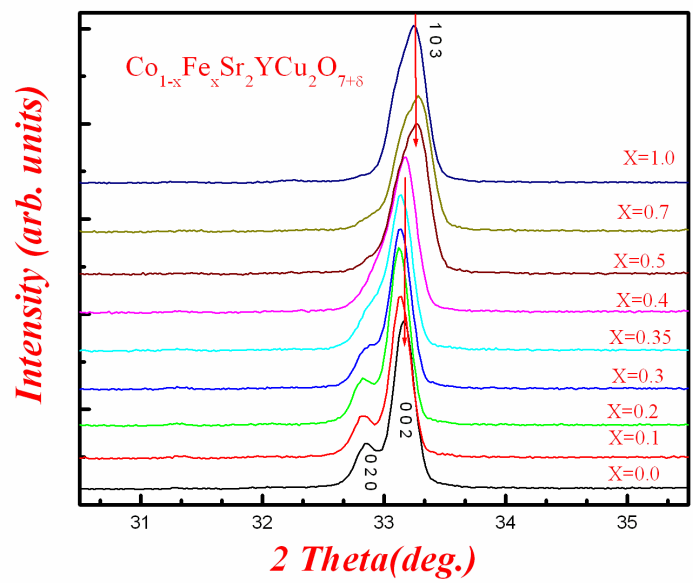

**Fig. 3a**

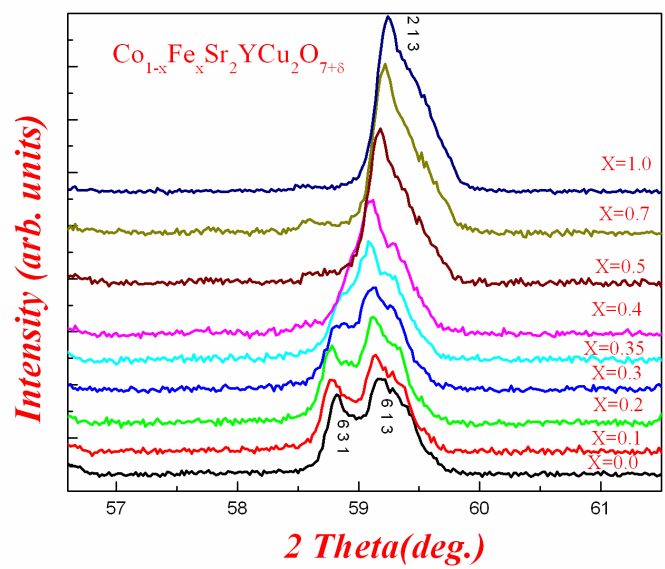

**Fig. 3b**

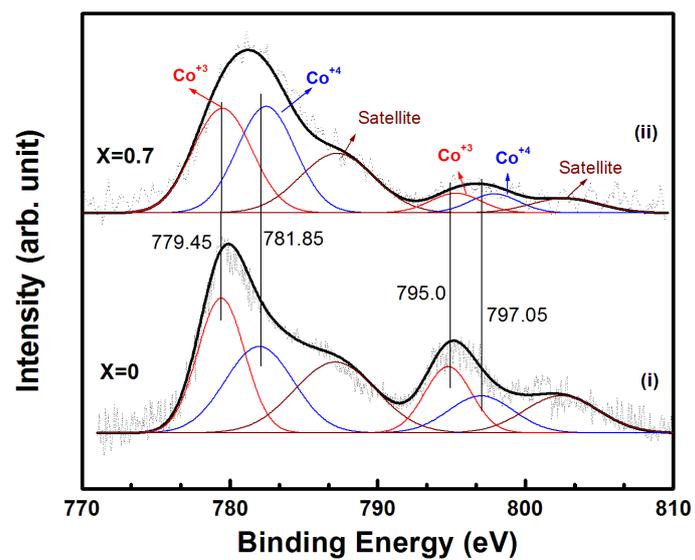

**Fig. 4a**

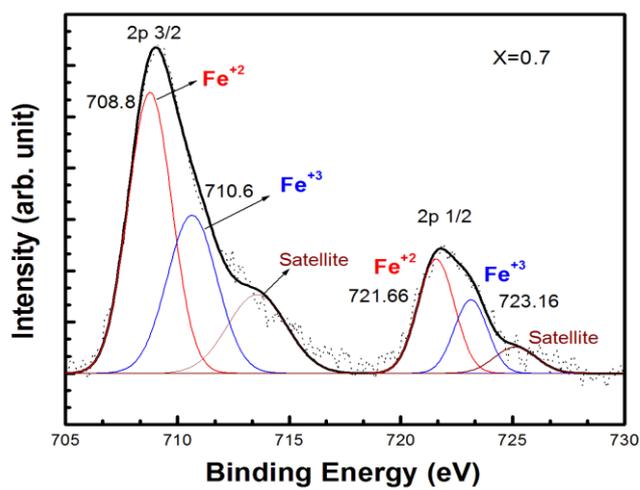

**Fig. 4b**

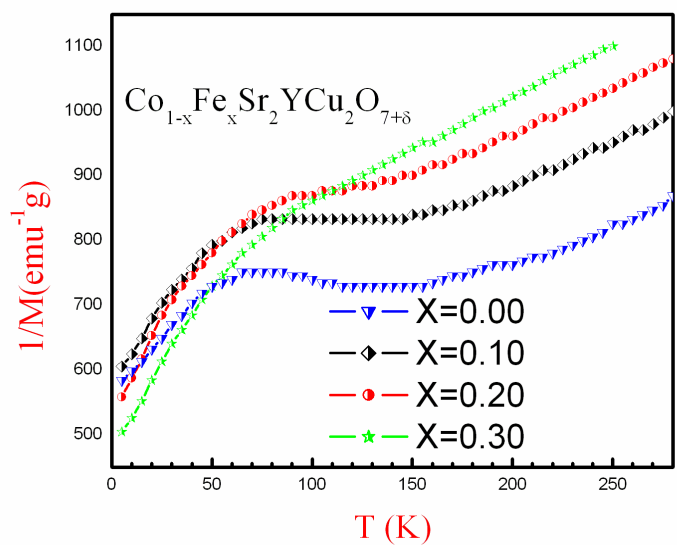

**Fig. 5a**

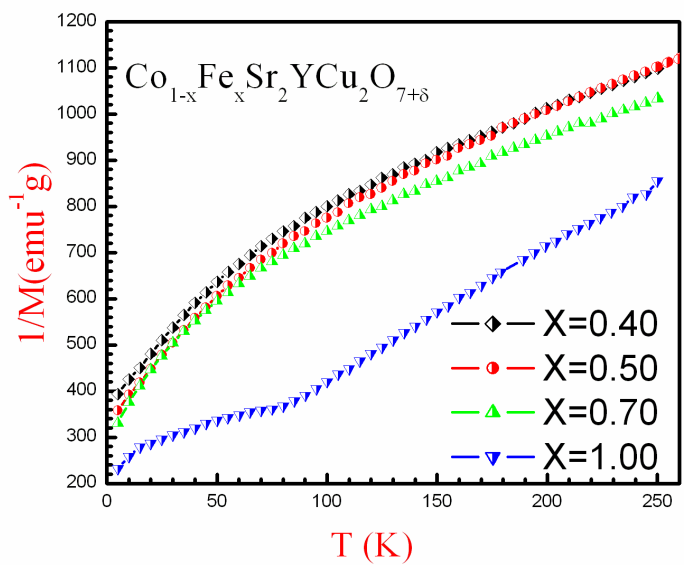

**Fig. 5b**

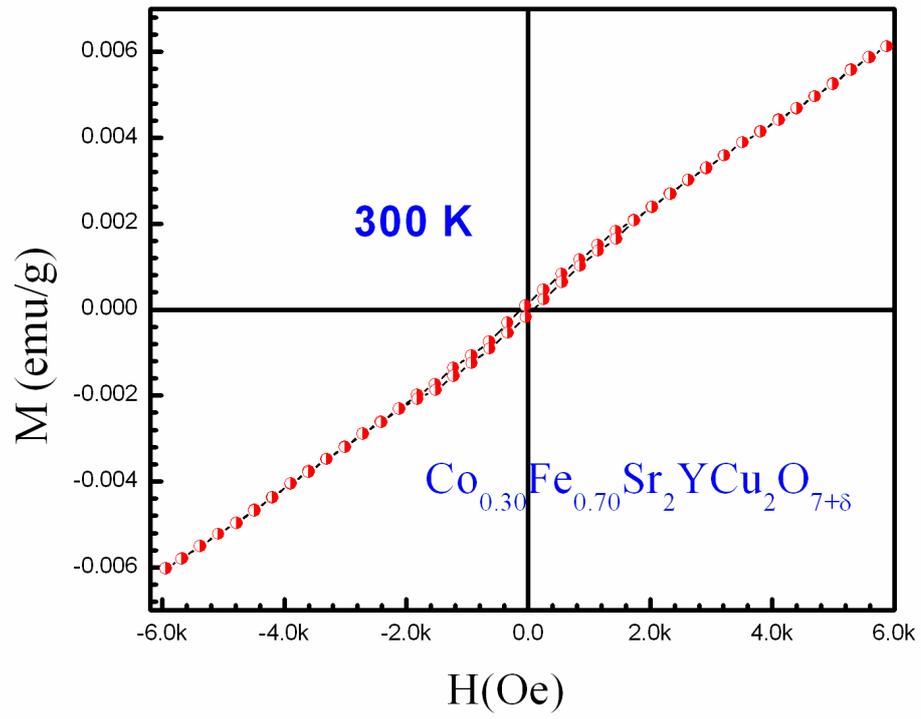

**Fig. 5c**

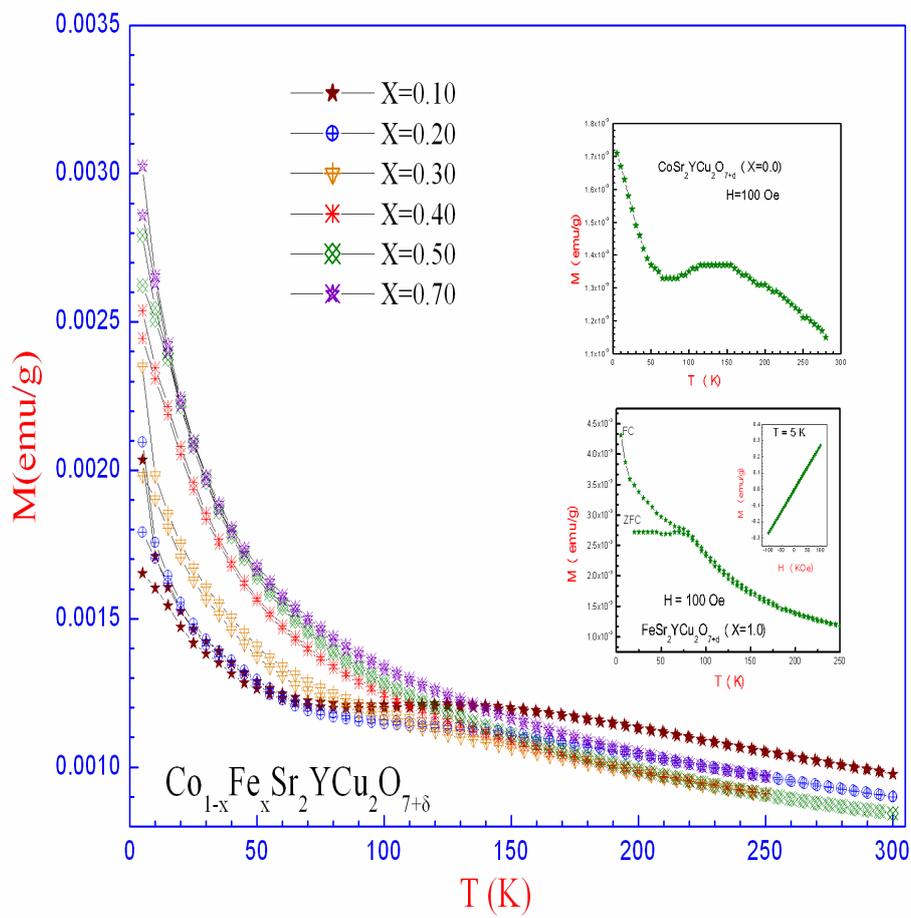

**Fig. 6**